\documentclass[conference]{IEEEtran}
\IEEEoverridecommandlockouts
\usepackage{cite}
\usepackage{amsmath,amssymb,amsfonts}
\usepackage{algorithmic}
\usepackage{graphicx}
\usepackage{textcomp}
\usepackage{xcolor}
\usepackage{caption}
\usepackage{subcaption}
\usepackage{tabularx,booktabs}
\usepackage{tabu}                      
\usepackage{float}

\def\BibTeX{{\rm B\kern-.05em{\sc i\kern-.025em b}\kern-.08em
    T\kern-.1667em\lower.7ex\hbox{E}\kern-.125emX}}
\begin{document}

\title{Predicting Human Performance in Vertical Hierarchical Menu Selection in Immersive AR Using Hand-gesture and Head-gaze\\}

\author{
    \IEEEauthorblockN{Majid Pourmemar $\dagger$, Yashas Joshi $\ddagger$, Charalambos Poullis}
    \IEEEauthorblockA{\textit{Immersive and Creative Technologies Lab, Department of Computer Science and Software Engineering}\\
    \textit{Concordia University}\\
    Montreal, Canada \\
    majid.pourmemar@concordia.ca}
}

\IEEEpubid{\begin{minipage}{0.5\textwidth}
 $\dagger$ Majid Pourmemar: conceptualization, methodology, software, data curation, investigation, validation, formal analysis, and writing of the manuscript. Corresponding author.
 $\ddagger$ Yashas Joshi: software.\\ \\ \\
\end{minipage}
\begin{minipage}{0.5\textwidth}
\hfill
\end{minipage}}



\twocolumn[{%
\renewcommand\twocolumn[1][]{#1}%
\maketitle
}]

\IEEEpubidadjcol
\begin{abstract}

There are currently limited guidelines on designing user interfaces (UI) for immersive augmented reality (AR) applications. Designers must reflect on their experience designing UI for desktop and mobile applications and conjecture how a UI will influence AR users' performance.
In this work, we introduce a predictive model for determining users' performance for a target UI without the subsequent involvement of participants in user studies. The model is trained on participants' responses to objective performance measures such as consumed endurance (CE) and pointing time (PT) using hierarchical drop-down menus. Large variability in the depth and context of the menus is ensured by randomly and dynamically creating the hierarchical drop-down menus and associated user tasks from words contained in the lexical database WordNet. Subjective performance bias is reduced by incorporating the users' non-verbal standard performance WAIS-IV during the model training. The semantic information of the menu is encoded using the Universal Sentence Encoder. We present the results of a user study that demonstrates that the proposed predictive model achieves high accuracy in predicting the CE on hierarchical menus of users with various cognitive abilities. To the best of our knowledge, this is the first work on predicting CE in designing UI for immersive AR applications. 
\end{abstract}
\IEEEpubidadjcol


\section{Introduction}

In recent years, Augmented Reality (AR) has emerged as one of the most promising technologies with applications in gaming, education, and medicine \cite{althoff2016influence},\cite{chang2019applying},\cite{pratt2018through}. Currently, AR hardware is categorized in terms of: (a) mobile AR, where the AR content is shown through hand-held tablets or smartphones, (b) spatial AR, which is based on projecting digital content on physical objects, and (c) immersive AR, where see-through head-mounted displays (HMDs) are used to augment the user's field-of-view. Each category of devices allows users to experience the real world overlaid with virtual, computer-generated content.
One of the most important characteristics of AR is the interactions with virtual content \cite{chatzopoulos2017mobile}. In immersive AR, this feature is supported through graphical User Interfaces (GUI) and input modalities relying on natural interactions such as hand-gestures, head-gaze, eye-gaze, and voice recognition. Thus, the design of GUIs and the choice of input modality is of paramount importance for immersive AR applications, and a determining factor for their usability. Moreover, in immersive AR, the interaction with the virtual content involves considerable physical or mental activity\cite{pourmemar2019visualizing},\cite{bach2018hologram}. Therefore, it is important to design these interactions in a way that reduces the amount of overall workload on the users. This will be more important for prolonged usage of an immersive headset, i.e. performing a sequence of interactions to reach a final goal. As the applications become more complicated, containing numerous actions and options, the popular choice amongst designers and developers is to combine the aforementioned menu options into groups and hierarchies(i.e. hierarchical drop-down menus as shown in Figure \ref{fig:teaser}).  

\begin{center}
    \centering
    \captionsetup{type=figure}
    \includegraphics[width=0.485\textwidth,height=5cm]{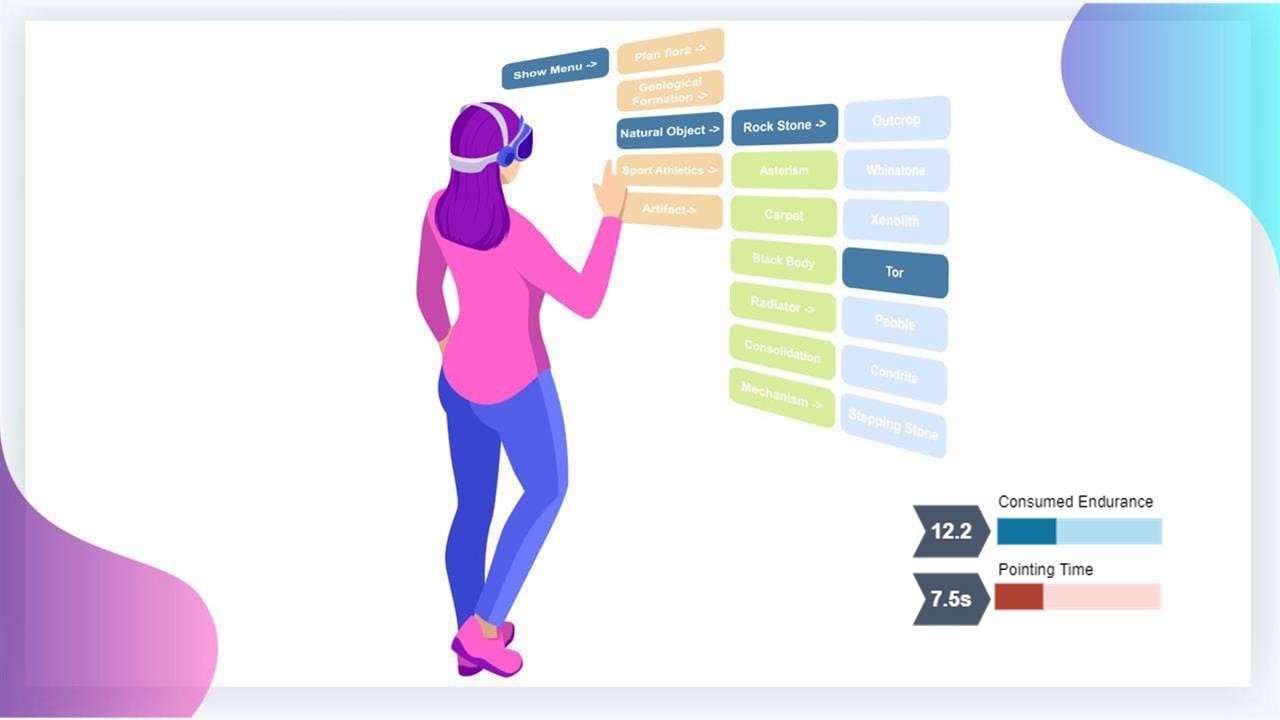}
    \captionof{figure}{We present an AI model for predicting human performance (pointing time and consumed endurance) in hierarchical menu selection in immersive Augmented Reality applications.}
    \label{fig:teaser}
\end{center}%

Although drop-down menus in immersive AR can be replaced with more user-friendly menus like radial \cite{gebhardt2013extended} or tile menus\cite{brudy2013interactive}, they are still widely used due to their familiar nature. This is especially the case for developers who transfer the two dimensional desktop menu layouts to three dimensional immersive AR. This means that optimizing the placement of the menu items in hierarchical drop-down menus can lead to improved interactions in immersive AR. To achieve this, one has to evaluate the performance of users of different cognitive capacities on the menu.

User studies must be conducted during and after the menu design process to provide insights into the usability concerning the effectiveness, efficiency, and overall satisfaction of the end-users. However, conducting user studies for immersive AR applications is a time-consuming and costly process. The participants must have access to an AR-capable HMD, e.g. HoloLens, Magic Leap, etc, which considering the current cost of HMDs is very unlikely. This restricts the type of user studies to on-site, face-to-face usability testing. This can be very time-consuming and expensive since participants are often compensated for their time. This is especially the case when an iterative design process is followed.

To overcome these challenges, we propose a predictive model that determines the users' performance for a target GUI without any involvement of participants during the user studies. The model is trained on data acquired during a user study involving a large number of hierarchical drop-down menus, and the responses of participants to objective performance measures.




\subsection{Human Performance on Hierarchical Menu Selection}


Hierarchical menus are widely used in a variety of applications ranging from simple web pages such as portals and websites, to highly professional desktop software like 3D authoring applications (i.e., Autodesk Maya and Unity3D). Each phase of the software contains hundreds of actions and options in a hierarchical menu format. According to the related literature, \cite{samp2010supporting, pourmemar2019visualizing} selecting a desired menu item, especially when it appears deep in the hierarchy, can be challenging. Based on several studies \cite{findlater2009ephemeral, li2018predicting}, a user's performance on menu selection in a desktop configuration, depends on the pointing time (PT). The PT is defined as the time it takes for the user to find the desired menu option and point to it, i.e. the time duration between the first and the last click. This is different for immersive AR/VR, since the input modalities for interacting with the GUIs are different. VR mainly uses controllers to interact with the GUI, while AR relies on head-gaze or hand-gestures. Studies specific to immersive AR/VR applications \cite{hincapie2014consumed, pourmemar2019visualizing} have shown that the user's performance for menu selection also depends on additional factors like workload and error rate. The workload is defined as physical or mental effort required to point the cursor at the desired menu item and select it. Since errors are more likely to occur in immersive AR than in the desktop environment\cite{bach2018hologram}, the error rate is also an important measure for evaluating the performance. Increased error rates are attributed to the fact that most interactions in immersive AR, are performed using hand-gestures and pointing the head-gaze. This increases the workload and amplifies the possibility of an error during a selection.


In this work, the user's performance on a hierarchical GUI is predicted considering the factors like expected PT and physical workload corresponding to selecting a menu item from a hierarchical drop-down menu through a sequence of menu selection tasks in immersive AR. Inspired by related work in the area, we consider the semantics and organization of the menu items (i.e. number of levels, number of items in each level, and number of characters for each target in the every level) and the users' cognitive ability in performing visual tasks (WAIS-IV test).








Previous works on performance prediction in HCI tasks categorize the measures into two sub-groups, user-related and GUI-related factors. A brief description of such factors is presented in this section.  

\subsubsection{User-related Factors}

The user-related factors depend on the capabilities of the users performing selection tasks. Their learning ability, cognitive performance, and fatigue are important factors that affect the performance of menu selection. These factors are recorded during the user study and considered in the final neural network for predicting the human performance. We applied a standard cognitive approach, WAIS-IV test \cite{wechsler2008wechsler}, to estimate each user's mind Processing Speed Index(PSI).

\subsubsection{GUI-related Factors}

The appearance of a menu is an essential factor in the user performance during a selection task. Shape of the menu (i.e., vertical list, radial) \cite{pourmemar2019visualizing}, saliency of items within, their semantic meaning, and the organization in general (i.e., number of levels)\cite{li2018predicting} are the most important factors. In this work, we focus on the vertical lists such as hierarchical menus with the maximum depth of 3 and consider the saliency, semantics, and organization of menu items as a part of the machine learning model input.

\subsection{Contributions}
The contributions of this work are as follows: 

\begin{itemize}

  \item  To the best of our knowledge, this paper is the first work on predicting human performance of vertical hierarchical menu selection in immersive AR incorporating both user-related and GUI-related factors.

  \item This work is the first work that considers the semantic relationship of the targets at different levels in the hierarchy. Previously, Li et al. \cite{li2018predicting}, focused only on the semantics of the words in vertical menus. We demonstrate that when it comes to a hierarchical menu selection, the semantic relationship between the targets in each menu level is an important factor to consider.
\end{itemize}

\section{Literature Review}
\label{Literature}

Human performance is an important factor in designing and developing user interfaces(UIs). The performance of a UI can be calculated using statistical methods and user studies with a certain number of participants\cite{browne2012empirical}. Although the performance of a UI can be evaluated using user studies, predicting the performance using extensive mathematical and machine learning models without testing it with real users will be less time-consuming and less expensive. The related work in predicting human performance in HCI is divided into two broad categories, namely the traditional methods that mostly use statistical and mathematical approaches such as Fitts' law\cite{fitts1954information} and Hick's law \cite{hick1952rate} and the novel machine learning methods such as \cite{li2018predicting, yuan2020modeling, swearngin2021modeling} which use deep neural networks to consider all user-related and GUI-related factors.


\subsection{Traditional Methods}

There are a plethora of mathematical models presented to predict human performance in HCI. Fitts' \cite{fitts1954information}, and Hick's law \cite{hick1952rate} are the two fundamental laws for predicting task completion time and form the basis for numerous prediction methods. The complex models such as \cite{bailly2014model} in the literature, predict visual search time based on Fitts' law. However, the model \cite{bailly2014model} predicts visual search time in vertical menus without considering the saliency of the items (i.e., number of characters in a word), which is one of the important factors in vertical menu selection. The work also does not consider the semantic meaning of each item as an important factor.


Considering the visual appearance of GUIs, the authors in \cite{jokinen2020adaptive} presented a computational model of visual search on graphical layouts. They gave several examples and user studies indicating how users learn GUIs and how changes in visual design will affect the performance of the UI. Bailly et al.\cite{bailly2014model} and Li et al. \cite{li2018predicting} call this measure as a learning effect and consider it in vertical menu selection.


Regarding the immersive environments, in \cite{erazo2015predicting}, the authors proposed a mathematical model for predicting task execution time for mid-air and touch-based human hand-gestures. Although, it predicts the task completion time with $R^2=0.936$ based on their own dataset gathered in a user study, they do not consider predicting the performance in a sequence of tasks with factors such as arm fatigue \cite{hincapie2014consumed}, learning effect \cite{li2018predicting} and GUI-related factors i.e. semantics and menu appearance.


\subsection{Machine Learning-based Methods}


Recent advances in machine learning and deep learning motivated HCI researchers to provide machine learning models for HCI applications. Therefore, predictive models of human performance are of significant interest to many researchers. Although traditional models perform very well in predicting human performance in simple cases with the condition of removing factors such as learning effect, semantics, task load and user's cognitive performance, they do not work well if all possible effective factors are considered. In such situations, the machine learning methods appeared to be resolving the issues\cite{yuan2021human}.


The authors in \cite{li2018predicting} presented a novel deep recurrent neural network to capture semantics, learning effect and saliency of menu items to predict visual search time in vertical menu selection. The method outperformed previous traditional works \cite{bailly2014model} in terms of accuracy and goodness of fit ($R^2$). However, the method does not explain the performance of hierarchical menu selection which is most common in many real use-cases. Another work that uses recurrent neural networks, Wei et al.\cite{wei2021predicting} presented a deep model to predict the mouse-click position in a sequence of interactions. The LSTM model was trained on previous paths of the mouse and could predict the future mouse-click area with a better accuracy and error rate compared to previous works including \cite{lank2007endpoint} and \cite{pasqual2014mouse}.

The aforementioned methods perform well in desktop environments. However, there is no indication that using the current 2D based models can also predict the performance in menu selection in 3D immersive environments using hand-gestures and head-gaze as the input modalities. In such environments the workload of the tasks is higher compared to desktop environments \cite{bach2018hologram} and it can affect the performance of the menu selections by user.

In conclusion, since in the immersive AR environment the possibility of noise and error and workload are higher, more complicated tools have to be applied to predict the performance of user. In the following sections, we present our method for predicting human performance in vertical hierarchical menu selection in immersive AR.

\vspace{-5pt}
\section{Methodology}
\label{Methodology}
\vspace{-5pt}

\begin{figure*}[t]
    \centering
    \includegraphics[width=.90\textwidth]{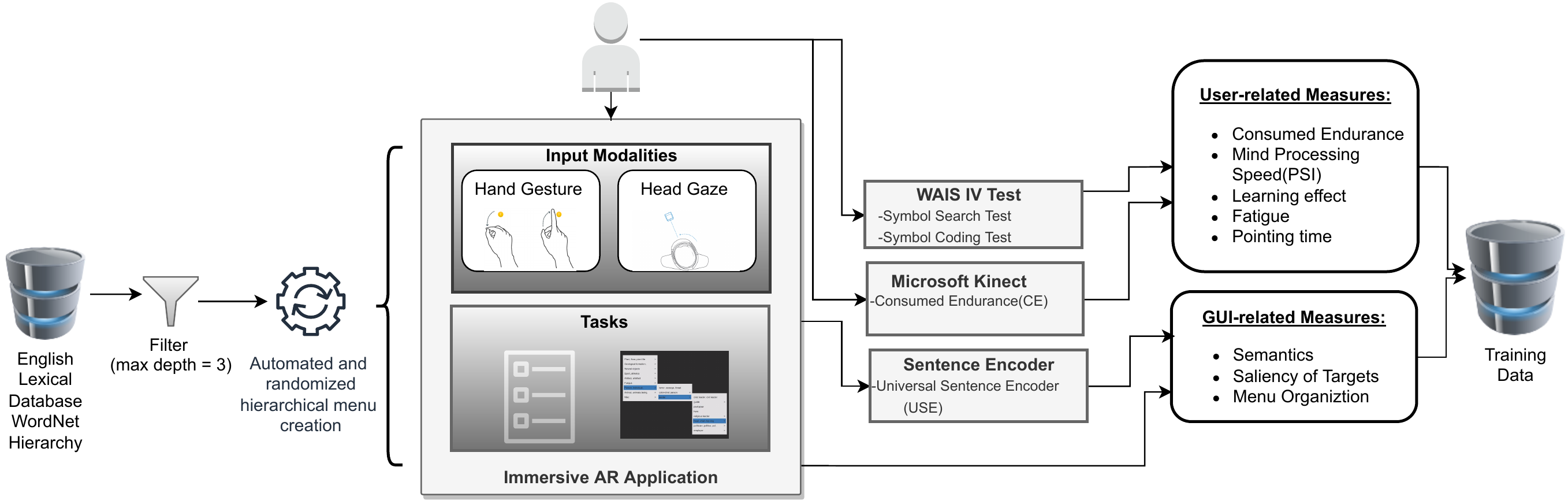}
    \caption{The diagram shows the data acquisition procedure during the user studies. An integrated AR application is developed on MS HoloLens I with hundreds of hierarchical drop-down menu selection tasks. The menu hierarchies were extracted from WordNet \cite{fellbaum2012wordnet}, a lexical database of semantic relations between the words in English. 
    The users performed the menu selection tasks through two separate user studies using hand-gestures and head-gaze with MS HoloLes I. A Microsoft Kinect I recorded the user's hand movements during the user studies. Finally, the data from the whole procedure is used to train and validate our machine learning model.
    }
    \label{fig:data_collection}
\end{figure*}

This research aims to provide a validated predictive model of CE and PT for hierarchical menu selection in immersive AR. Two user studies were conducted to gather the required training data, and evaluate the machine learning model. The proposed model was trained on the data from the first user study and evaluated through a set of unseen data collected from the second user study. In the following sections, we present details for these user studies. The diagram in Figure \ref{fig:data_collection} summarizes the components of data collection procedure during the user studies.  

\subsection{Application}
\label{subsec:application}

We developed an immersive AR application on MS HoloLens I, coupled with a Microsoft Kinect I for data acquisition. The integrated application is developed with Unity3D and includes two separate modes for training and evaluation. Microsoft's Mixed Reality Toolkit (MRTK) is employed to develop immersive AR features and interactions. The applied input modality was the combination of hand-gestures (for menu selection) and head-gaze (for moving the cursor), as they are most common in immersive AR headsets. The GUI consists of a regular hierarchical drop-down menu as this is widely used in AR/VR applications. The hierarchical menu contains hundreds of drop-down menu selections for different tasks in an immersive AR environment. Since the usual menu depths in any typical software varies between 2 to 3 (i.e., in 3D authoring applications such as Autodesk Maya or 3D Studio Max, or Unity3D), we have selected 3 as the maximum menu depth in our application. 
Each user selection is randomized and the task depth varies from 2 to 3. A display prompt on the top showed the item to be selected from the hierarchy, and the users were instructed to perform that selection using either hand-gestures or head-gaze. The PT and user's hand movements for each selection were recorded using the built-in application's procedure with Microsoft Kinect. In order to have a complete hierarchy of words in our menu, we used the taxonomy of WordNet \cite{miller1995wordnet}, which is an extensive lexical database of English nouns, verbs, and adjectives. For example, "sport $\rightarrow$ cycling $\rightarrow$ dune cycling" is showed on the display prompt, asking the user to select dune cycling from the hierarchy, as a single task. The semantics of the words on each level and the relation between the words are applied in the performance prediction pipeline.

\subsection{Measures}
\label{subsec:measures}

In addition to measuring the pointing time (PT) for each menu selection task, we also measure CE for evaluating physical workload, WAIS-IV test for assessing the participants' cognitive performance, Semantics, and Organization of menus for assessing the GUI-related aspects of menu selection. This section discusses these measures.

\subsubsection{Consumed Endurance (CE)}

The authors in \cite{hincapie2014consumed} proposed a measure for evaluating workload in tasks with mid-air hand-gestures. This method can be used in immersive AR/VR applications that rely on mid-air hand-gestures as input modalities. CE requires a Microsoft Kinect to track the user's hand movements in real-time and assigns a workload value to the 3D manipulation task. The model is derived from the biomechanical structure of the upper arm for both male and female users, and it is widely used as a metric for evaluating the Gorilla Arm effect. Our proposed model predicts CE for hierarchical menu selection tasks.
 
\subsubsection{WAIS-IV Test}
Prior to performing the menu selection tasks, the participants were asked to complete a paper-based subtest of WAIS-IV \cite{wechsler1958measurement} to measure their cognitive ability to perform visual menu tasks. This data lets the machine learning model discriminate between users in performing menu selection tasks and provide the designers and developers with insights on their design's personalized performance. The complete WAIS-IV test is a standard test that covers all aspects of human intelligence, including Verbal IQ (VIQ) and Performance IQ (PIQ). For our experiments, the Processing Speed Index(PSI) is measured as part of the PIQ. The PSI test consists of two subtests: Symbol Search and Symbol Coding. The symbol search test includes 63 symbol search matching questions printed on seven one-sided A4 papers. The users were asked to find similar symbols in each line and mark them during the standard 120 seconds to complete the test.
On the symbol coding test, the user had 120 seconds to copy symbols paired with the numbers. 
The output will be two integer numbers, symbol search results between 0 and 63 and symbol coding between 0 and 135.


\subsubsection{Semantics and Menu organization}

Based on the results of \cite{li2018predicting} presented by Li et al., the semantics of menus affect the performance of vertical menu selection. Therefore, including the semantics of menu items forming a list improves the accuracy of the predictions. In our work, we also extract the semantics of the menu items using word embedding techniques. A novel encoding approach named Universal Sentence Encoder (USE) \cite{cer-etal-2018-universal} is applied on the menu items to encode the varied length menu lists to a fixed length one. USE takes as input a sentence and produces a 512-dimensional vector representing the items' semantic meaning.


The hierarchical menu organization is also considered as one of the inputs for the machine learning model. For each target in each task, the menu item's position in the list coupled with the number of whole menu items in that list is considered. The saliency of the targets is also considered and is calculated as the number of characters for each word.

\subsection{User Study Procedure}
\label{Procedur}

\begin{figure*}[ht]
    \begin{center}
    \centering
    \includegraphics[width=.90\textwidth]{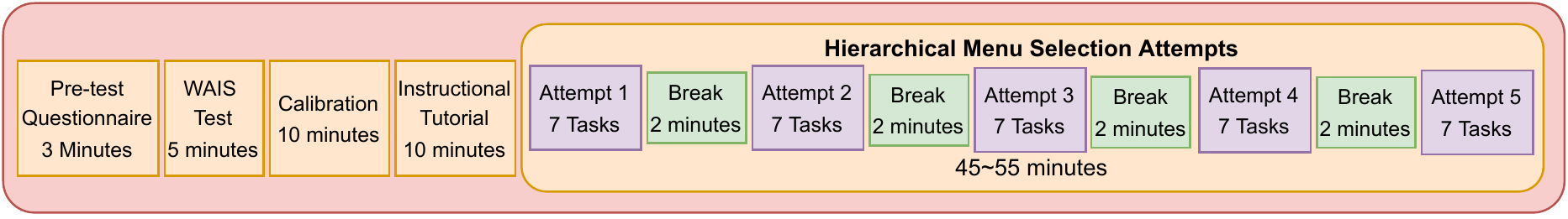}
    \end{center}
    \caption{
    The participants had to fill out a pre-test questionnaire followed by a 5 minutes standard WAIS-IV Test and a 10-minute calibration to adjust the headset. A 10-minute instructional tutorial to familiarize the participants with hand-gestures used in immersive AR, e.g., air-tapping, was presented as the next step. Then, the participants had to do the main part of the test: the menu selection tasks. The section consists of five menu selection attempts, including seven hierarchical menu selection tasks. In order to avoid users' fatigue, each attempt was followed by a 2-minute break.
    }
    \label{fig:procedure}
\end{figure*}

\begin{figure}
    \centering
    \includegraphics[width=0.34\textwidth]{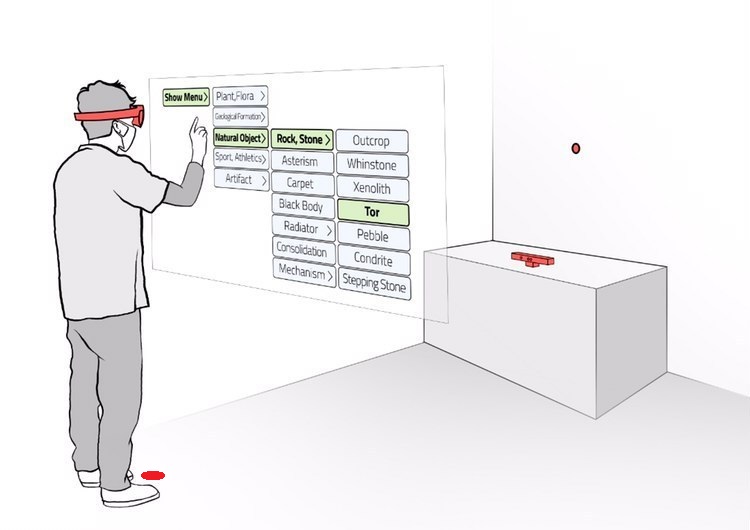}
    \caption{The participant is performing menu selection tasks while the MS Kinect records the hand motions. The red dots on the ground and on the wall in front of user are used for having similar initiation configuration for all users. 
    }
    \label{fig:experiment}
\end{figure}

This research received approval from the University's Research Ethics Committee. Two user studies were conducted to gather training data and evaluate the machine learning model. The participants were assigned through a random process to do either the first or second user study. The participants were first asked to complete a pre-test questionnaire containing questions relating to their demographics and background experience in using immersive AR technologies through each user study. Then they were asked to complete the WAIS-IV test in the researcher's presence.
A 10-minute calibration of HoloLens I and a 10-minute instructional tutorial were the following steps before starting the main hierarchical menu selection attempts. Participants were asked to perform menu selection tasks with MS HoloLens I during the main test. This part included five menu selection attempts consisting of seven menu selection tasks. The number of attempts and tasks resulted from a pilot study conducted before the main study considering the user's subjective feedback of the test. Considering the high physical demand of immersive AR tasks using hand-gestures, the users were given a 2-minute break between performing each menu selection attempt. The menu selection tasks were presented in a single line of instruction at the top-right corner in the user's view, informing the user which menu item should be selected. A Microsoft Kinect recorded the participant's hand movements during the activity, and the CE was calculated per task. Figure \ref{fig:procedure} shows the procedure followed and Figure \ref{fig:experiment} shows the setup of environment during the user studies.

\subsection{Participant Demographics}
\label{subsec:participants}
A total number of 26 users (12 female, 14 male) participated in our user study for both training and evaluation parts. The participants were randomly assigned to the training/evaluation user study. The participants ranged from 24-39 years old with the majority of 73\% within the age group 27 to 35 and 11\% within group of 35 to 39 and 16\% within the age group of 23 to 27.


\vspace{-10pt}
\section{Model Design and Learning}
\vspace{-5pt}

\label{sec:model_design}

We trained a recurrent neural network  on user-related and GUI-related measures extracted from the user studies to predict the PT and CE of hierarchical menu selection tasks in a sequence of interactions. The input to the model consists of three parts, including normalized WAIS-IV test data representing the user-related measures, organization of menu items, and semantic vectors as it is shown in the Figure \ref{fig:system_overview}. The model's output is the desired CE and PT through a sequence of tasks. We took advantage of a recurrent neural network to assess the user's activity through a series of tasks and consider the effects of the learning effect and the user's fatigue while performing the tasks. In this section, first, we will explain the data preprocessing for preparing the data for feeding to the machine learning model, Section \ref{subsec:preprocessing}, second the architecture of the machine learning model in Section \ref{subsec:model} and finally, the training of the model in Section \ref{subsec:training}.


\subsection{Data Prepossessing}
\label{subsec:preprocessing}
\begin{figure}[t]
  \centering
  \begin{subfigure}[t]{0.24\textwidth}
        \centering
         \includegraphics[width=\textwidth]{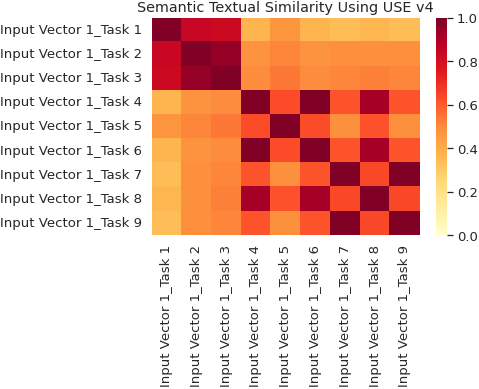}
         \caption{Semantic similarity map for input vector 1 for 9 selected tasks}
         \label{fig:InputVector1}
  \end{subfigure}
    \hfill
    \begin{subfigure}[t]{0.24\textwidth}
        \centering
         \includegraphics[width=\textwidth]{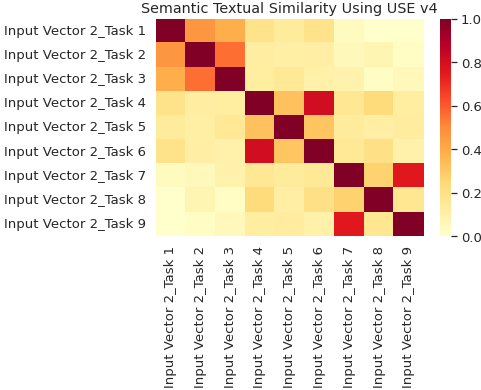}
         \caption{Semantic similarity map for input vector 2 for 9 selected tasks}
         \label{fig:InputVector2}
  \end{subfigure}
  \hfill
      \begin{subfigure}[t]{0.24\textwidth}
        \centering
         \includegraphics[width=\textwidth]{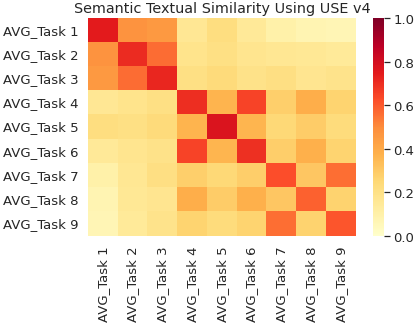}
         \caption{Semantic similarity map for average value of input vectors 1 \& 2}
         \label{fig:SemanticAVG}
  \end{subfigure}
    \hfill
      \begin{subfigure}[b]{0.35\textwidth}
        \centering
         \includegraphics[width=\textwidth]{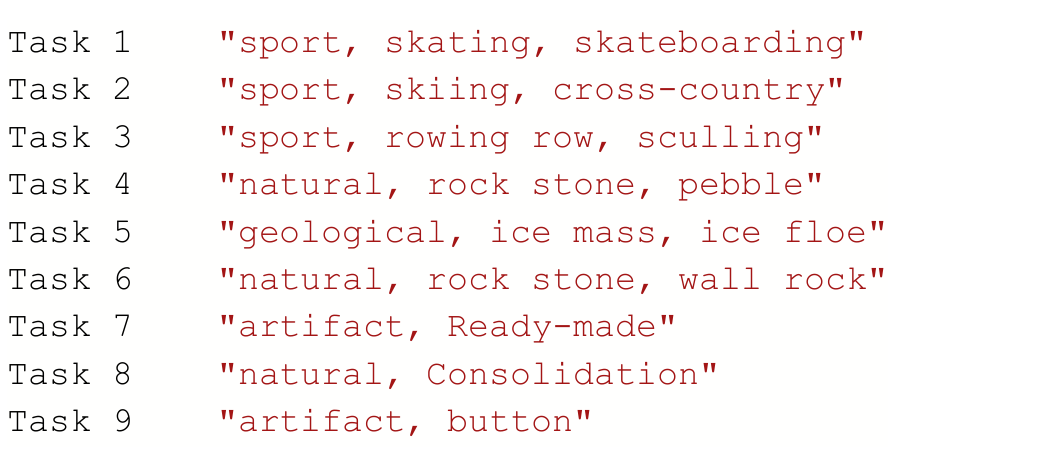}
         \caption{The targets for each menu selection tasks}
         \label{fig:Tasks}
  \end{subfigure}

  \caption{ USE \cite{cer2018universal} is applied on the tasks to capture the semantic similarity by converting it to a fixed-length vector. The colors along with the diagonal show semantic similarity between the vectors encoding the tasks.
  }
  \label{fig:USETaskSimilaity}
\end{figure}

Menus and hierarchies have variable length and depth. To address this problem, we converted the sequence of menu items to fixed-length vectors using word embedding techniques as discussed in \cite{li2018predicting} and employed the USE technique \cite{cer2018universal}. The preprocessing was applied to the inputs as it is indicated in Figure \ref{fig:system_overview}. Since the USE receives sequences of words, we created comma-separated sentences from menu items, including all menus from the start to the end of the list, namely Input Vector 1. Another sentence was also produced from targets in each level, separated by a single comma to emphasize each level's target, emphasizing semantic relation between the targets, namely Input Vector 2. The word vectors were later fed to USE to be converted to fixed-length vectors. Figure \ref{fig:USETaskSimilaity} shows an example of the similarity of the fixed-length embedding vectors for nine different tasks.


    

\subsection{Neural Network Model}
\label{subsec:model}

\begin{figure*}[!ht]
    \centering
    \includegraphics[width=.90\textwidth]{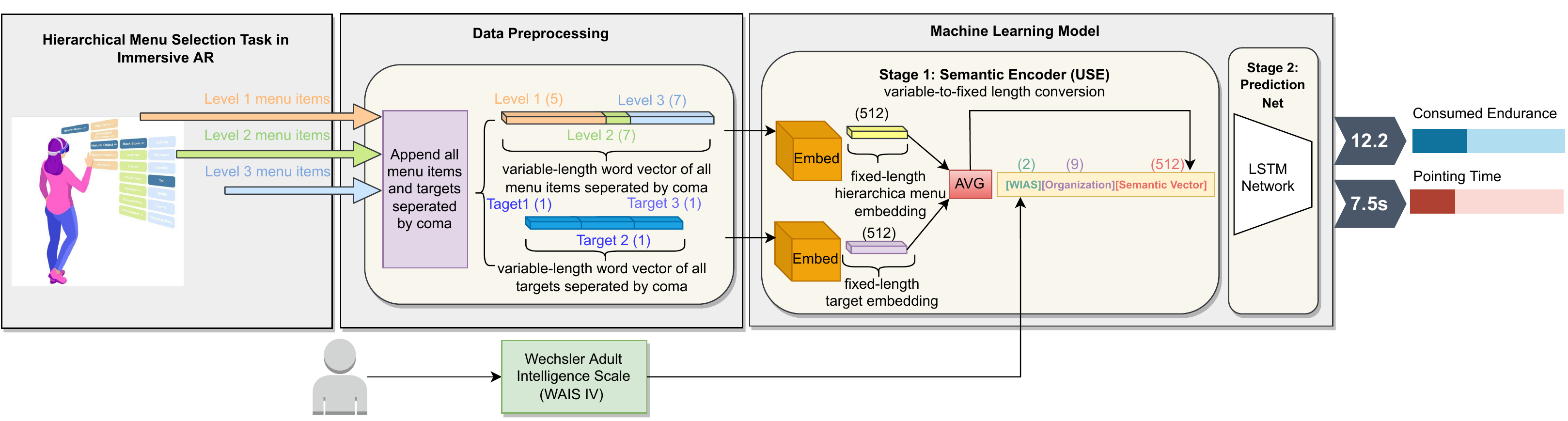}
    \caption{
    The data preprocessing and machine learning pipeline. In order to convert varied length menu items to the fixed length vectors firstly, a sequence of all items starting from the beginning of the first-level to the end of last level d (d=2, 3) separated by single comma between each two words created, Input Vector 1. Secondly a sequence consisting of all target items in each level, separated by single comma is created, Input Vector 2. The two sequences were passed to the USE network \cite{cer2018universal} to be converted to a fixed-length 512-dimensional vectors. We applied the pre-trained models of USE which is available on TensorFlow Hub . Finally, we found that the element-wise average of two vectors lead to better results for the validation set. The WAIS-IV data and the organization of the tasks are also concatenated to the result of the semantic vector.
    }
    \label{fig:system_overview}
\end{figure*}

The machine learning model consists of the encoder and the prediction networks, as shown in Figure \ref{fig:system_overview}. To encode the varied length hierarchies, we used USE \cite{cer2018universal} which is a transformer-based model designed to convert the varied-length paragraph to fixed-length float value vectors. After converting the tasks to fixed-length vectors, the prediction network is trained on the user's performance on a sequence of tasks to predict the PT and CE. 
We used an LSTM model with one hidden layer, an input layer size of 523, and an output layer size of 2 (CE and PT).
The training window size for considering the previous tasks was 15, meaning that the prediction of performance for each menu selection task is based on the previous 15 tasks performed by each user. Therefore, the recurrent model was able to predict the learning effect through the tasks.

\subsection{Training}
\label{subsec:training}
We trained our model on a workstation with NVIDIA Geforce RTX 2060 GPU using the PyTorch library. The training data size is 840, and the test data size is 140, including the menu selection tasks done by users. Since this is a regression problem, we optimized the Mean Squared Error (MSE) during the training. This is equivalent to optimizing the $R^2$, which is a measure for determining the goodness of fit. We use the Adam optimizer with a learning rate of 0.00002. In addition, we found that the dropout rate of 0.3 reduces over-fitting.

\vspace{-5pt}
\section{Discussion and Analysis of the Results}
\vspace{-5pt}

\label{sec:experimental_results}

\begin{figure}[ht]
     \centering
     \begin{subfigure}[b]{0.34\textwidth}
         \centering
         \includegraphics[width=\textwidth]{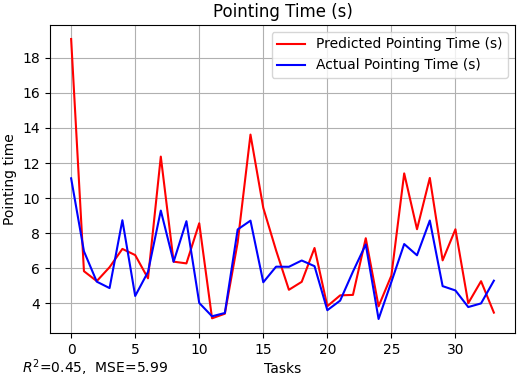}
         \caption{}
         \label{fig:pt1}
     \end{subfigure}
             \begin{subfigure}[b]{0.34\textwidth}
         \centering
         \includegraphics[width=\textwidth]{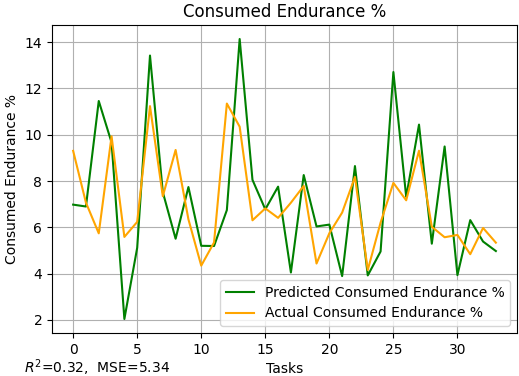}
         \caption{}
         \label{fig:ce1}
     \end{subfigure}
        \caption{The predicted CE/PT versus the actual CE/PT through 35 tasks done by a participant in the evaluation user study. The vertical axis is the CE/PT and the horizontal axis represents the order of task from 0 to 34 (as it is depicted in Figure \ref{fig:procedure} each participant performed 35 tasks in 5 attempts each one included 7 tasks). }
        \label{fig:PredictionsCE}
\end{figure}

 We have evaluated our model using test data extracted from the evaluation user study. The model was trained on optimizing $R^2$ as the goodness of fit. In other words, the aim is to regress the PT and CE for a sequence of unseen tasks from the test data gathered from the user study with four random participants having presented to them completely different menu hierarchies than the ones in the training data. Therefore, the model was trained to capture different aspects that affect performance including semantics, menu organization and users' cognitive performance. The Figure  \ref{fig:PredictionsCE} shows the predictions and actual values for PT and CE for one of  the users from the evaluation user study. Table \ref{ResultsCE} shows the results of the predictions for each of the users who participated in the evaluation user study. Each table row consists of the $R^2$ and MSE for each user. The results show that our model effectively predicts the CE and PT in a sequence of hierarchical menu tasks in immersive AR. Our investigations also show that the model performs better after learning the users behaviors in selecting the menu options i.e after a number of tasks performed by the user. We confirmed this by evaluating the average of error for the first five tasks and last five tasks of each user.
 
 Although we achieved lower $R^2$s compared to the previous works such as \cite{li2018predicting}, our work has some specifications that make it different from other works. Firstly we trained our model on the data gathered in an immersive AR application. Interacting with virtual content in an AR application includes a lot of inconsistency and noise since the input modalities are head-gaze and hand-gesture as the work \cite{pourmemar2019visualizing} shows that the error rate in menu selection in such an environment is high. On top of that, most of the users who participated in our user study had almost no previous experience in working with immersive AR/VR headsets. As we asked from users in the pretest questionnaire, only 7\% of them claimed they had good experience using AR headsets. That was why we added an instructional tutorial for each participant before starting the menu selection tasks. Secondly, in this work, we employ hierarchical menu designs instead of a simple drop-down menu with only one depth level. Having a hierarchical task may increase the chance of error and inconsistency by itself. Thirdly, our experiments on the test data for immersive AR tasks, shows a positive correlation between the accuracy of the model and increasing the size of the training data.

\begin{table}
\caption{Results of Average $R^2$ and MSE for predictions PT and CE on the evaluation dataset }
\label{ResultsCE}
\begin{tabular}{  m{1.38cm}  m{1.38cm} m{1.38cm}   m{1.30cm} m{1.30cm} } 
    \toprule
    User  & $R^2$ CE&MSE CE& $R^2$ PT &MSE PT \\
    \midrule
    1   &   0.32    &   5.34    &   0.45    &   5.99 \\
    2   &   0.08    &   9.92    &   0.28    &   7.03\\
    3   &   0.31    &   7.46    &   0.4     &   8.17\\
    4   &   0.26    &   6.06    &   0.31    &   6.37 \\
Average &   0.24    &   7.19    &   0.36    &   6.89\\
    \bottomrule
\end{tabular}

\end{table}

\subsection{Ablation}
In order to evaluate the effect of each input component on the predictions, an ablation study is conducted by removing from the input vector the WAIS-IV data, and semantic and organization of menus. The results of training the network without these components are shown in Table \ref{Ablation}. The ablation study shows that the combination of semantics, WAIS, and the organization of menus appeared to be accounts for much of the increase in the accuracy of prediction.

\begin{table}
\caption{Results of Ablation study to determine the importance of each feature}
\label{Ablation}
\begin{tabular}{  m{2.48cm}  m{1.09cm} m{1.09cm}   m{1cm} m{1cm} }
    \toprule
Feature to Remove  &  $R^2$ CE&  MSE CE &$R^2$ PT & MSE PT \\
\midrule
No Remove               &   0.24    &   7.19    &   0.36    &   6.89    \\
Without WAIS            &   0.11    &   11.23   &   0.25    &   8.50    \\
Without Semantics       &   0.03    &   8.74    &   0.14    &   7.11    \\
Without Organization    &   0.0     &   15.67   &   0.1     &   12.37   \\

\bottomrule
\end{tabular}
\end{table}

The analysis of the results also shows a positive correlation between the WAIS-IV test results and the performance of users in immersive AR, and thus integrating the WAIS results for a user improves the accuracy of the prediction. This also confirms the reported correlation between the WAIS-IV test result and the PT and amount of physical activity for each task. A Pearson's correlation analysis has been applied on the results of the WAIS-IV (Symbol Search and Symbol coding) for different participants and their CE measurements, PT and error rate. The results clearly showed that there is a moderate degree of negative correlation between the WAIS-IV results and the CE values as well as PT. The Table \ref{table:WAISCorrelations} shows the correlation between PT, CE, WAIS data, and the age of the participants.


\begin{table}

\caption{Pearson's correlation coefficient, $\rho$, indicating the correlation between the WAIS-IV data and our
performance measures}
\label{table:WAISCorrelations}
\begin{tabular}{  m{2.22cm}  m{1.18cm} m{1.18cm}   m{1.10cm} m{1.10cm} }
\toprule
Performance Factor  &  Total WAIS &  Symbol Search & Symbol Coding  & Age \\
\midrule
CE  &   -0.324  &-0.443     &-0.167     &-0.11\\
PT  &   -0.284  &-0.295     &-0.214     &-0.434 \\

\bottomrule
\end{tabular}

\end{table}


\vspace{-10pt}
\section{Conclusion and Future Work}
\vspace{-5pt}
Predicting human performance is beneficial in designing and developing user interfaces. Various works currently predict human performance in desktop or smartphones that works perfectly but not any recent work in immersive AR. Interacting with virtual content in immersive environments is mainly done by natural interactions such as hand-gestures and head-gaze. These natural interactions can increase the possibility of error, noise, and user fatigue. In this work, we presented a method for predicting human performance of hierarchical menu selection in immersive AR using hand-gestures and head-gaze as the input modalities. Our model builds upon previous work in the area for vertical menu selection in desktop environments \cite{li2018predicting}. The features involved in the performance of menu selection are divided into two categories: user-related and GUI-related measures. We have applied both categories to reach better results for predicting human performance through a sequence of tasks. To this end, we conducted two user studies to gather the data used for training and evaluating the machine learning model. The results show that applying the cognitive WAIS-IV test improves the predictions and reduces the effect of environmental noise and users' fatigue. A natural extension of this work is its application to other graphical user interfaces such as radial menus and input modalities such as hand-gestures and head-gaze.


\vspace{-5pt}
\section*{Acknowledgements}
\vspace{-5pt}
This research is based upon work supported by the Natural Sciences and Engineering Research Council of Canada Grants No. N01670 (Discovery Grant).

\vspace{-10pt}

\bibliographystyle{unsrt}
\bibliography{HSI2022_Prediction}

\end{document}